\documentclass[%
 reprint,
superscriptaddress,
 amsmath,amssymb,
 aps,
]{revtex4-2}

\usepackage{amsmath,amssymb,amsfonts,graphicx,url} 
\graphicspath{{./},{./figures/}}

\newcommand{\ud}{d} 

\newcommand{\op}[1]{\operatorname{#1}}

\newcommand{\trace}{\op{Tr}}

\usepackage{graphicx}
\usepackage{epstopdf}
\usepackage{graphicx}
\usepackage{dcolumn}
\usepackage{bm}

\begin{document}

\preprint{APS/123-QED}

\title{Thermodynamic engine powered by anisotropic fluctuations}
\thanks{All authors contributed equally to this work. Numerical experimentation was carried out by OM with assist from AT. The conceptual frame was conceived collaboratively by all.}%
\author{Olga Movilla Miangolarra}
\affiliation{%
Department of Mechanical and Aerospace Engineering, University of California, Irvine, CA 92697, USA
}
\author{Amirhossein Taghvaei}
\affiliation{%
Aeronautics and Astronautics  Department, University of Washington, Seattle, WA 98195, USA
}%
\author{Yongxin Chen}
\affiliation{%
School of Aerospace Engineering, Georgia Institute of Technology, Atlanta, GA 30332, USA
}%
\author{Tryphon T. Georgiou}
\affiliation{%
Department of Mechanical and Aerospace Engineering, University of California, Irvine, CA 92697, USA
}%

\date{\today}

\begin{abstract}
The purpose of this work is to present the concept of an autonomous Stirling-like engine powered by anisotropy of thermodynamic fluctuations. Specifically,  simultaneous contact of a thermodynamic system with two heat baths  along coupled degrees of freedom generates torque and circulatory currents -- an arrangement referred to as a Brownian gyrator.
The embodiment that constitutes the engine includes an inertial wheel to sustain rotary motion and average out the generated fluctuating torque, ultimately delivering power to an external load. We detail an electrical model for such an engine that consists of two resistors in  different temperatures and three reactive elements in the form of variable capacitors. The resistors generate Johnson-Nyquist current fluctuations that power the engine, while the capacitors generate driving forces via a coupling of their dielectric material with the inertial wheel. A proof-of-concept is established via stability analysis to ensure the existence of a stable periodic orbit generating sustained power output. We conclude by drawing a connection to the dynamics of a damped pendulum with constant torque and to those of a macroscopic Stirling engine. The sought insights aim at nano-engines and biological processes that are similarly powered by anisotropy in temperature and chemical potentials.
\end{abstract}
\maketitle

\section{Introduction}
Carnot's 1824  abstraction of a heat engine \cite{carnot1824reflections} led to the discovery of entropy and to the birth of thermodynamics. In the intervening two centuries, in spite of great strides, very many conundrums lingered, largely due to the absence of models that capture the time-scale separation of processes involved. Today, we are witnessing the emergence of finite-time thermodynamics \cite{Jarz1996eq,Crooks1999FT,Evans1994FT} and of stochastic models \cite{seifert2012stochastic,sekimoto2010stochastic} that have brought about a finer understanding of those issues that were raised by Maxwell, Boltzmann, Loschmidt, and other founders of the field long ago.

The present work explores the coupling between (fast) thermal fluctuations and a (slow) mechanical component in a way that allows generation of mechanical power. 
Inspired by nature's ability to harvest energy from fluctuations and anisotropic chemical concentrations in conjunction with varying electrochemical potentials \cite{battle2016broken,gnesotto2018broken}, we introduce
an ``engine concept'' based on the model of a Brownian gyrator $-$ a system with two degrees of freedom that exhibits a characteristic non-equilibrium steady-state circulating current due an anisotropic temperature field.

Previous work on the Brownian gyrator focused on the circulating current and torque generated at steady-state \cite{BGyrator2007first,BGyrator2013CilibertoExperim,BGyrator2013dotsenko,BGyrator2017electrical,BGyrator2017experimental,Imparato2017BG,BGyrator2017electrical,gyratingcharacnonharmonic} -- experimentally validated in \cite{BGyrator2017experimental,BGyrator2017electrical,BGyrator2013CilibertoExperim,Bgyrator2013ciliberto}, on optimal transitioning between states  \cite{baldassarri2020engineered}, and the role of information flow  \cite{BGinfo,Loos_2020nonreciprocal}. Other works focused on underdamped mesoscopic systems \cite{inertiaBG}, non-Markovian noise \cite{NonmarkovianBG}, active reservoirs \cite{activereservoirBG}, and the effect of external forces \cite{constantforceBG} on such two-dimensional stochastic systems.
More recently, \cite{EnergyHarvestingAnisotropic2021}  considered the energetics of the cyclic operation of Brownian gyrators and derived theoretical bounds on efficiency and power that can be extracted from the anisotropy of the temperature field.

Our embodiment of the Brownian gyrator, following \cite{BGyrator2017electrical,Bgyrator2013ciliberto,BGyrator2013CilibertoExperim}, consists of a simple electrical network composed of two resistors
and three capacitors. Johnson-Nyquist fluctuating currents at the two resistors,
due to uneven ambient temperatures, allow for the potential to generate torque.
We postulate variable capacitors with moveable dielectric material. Forces
on the dielectric material are exerted by the fluctuating currents as well as by a
coupling to a flywheel. This mechanical component provides inertia and dissipation that absorbs generated power.  We provide detailed analysis as a proof-of-concept for the feasibility of this power generating mechanism,
and explain the mechanics responsible for energy transfer between the fluctuating currents and the rotational subcomponent of the engine.

Proposals for heat engines that are powered by thermal excitations date back to Maxwell's demon \cite{bennett1982thermodynamics,strasberg2013thermodynamics} and the Feynman ratchet \cite{parrondo1996criticism}. Experiments to validate relevant thought experiments
have been reported recently \cite{toyabe2010experimental,berut2012experimental,Martinez2015exp,bang2018experimental,ciliberto2017experiments}. However, these experimental demonstrations are mostly based on manipulating particles  in a non-autonomous manner, via optical traps and externally specified cyclic control protocols \cite{mechAutonomousheateng,bryant2020energy}. Such nano-manipulation requires considerable energy that far exceeds work that is being produced. The point of our paper is to present an analysis for the coupling between the system responsible for thermal fluctuation with a slower mechanical component that renders the operation of the engine autonomous.


\section{Brownian gyrator} 
\begin{figure}
\centering
\includegraphics[width=0.8\linewidth]{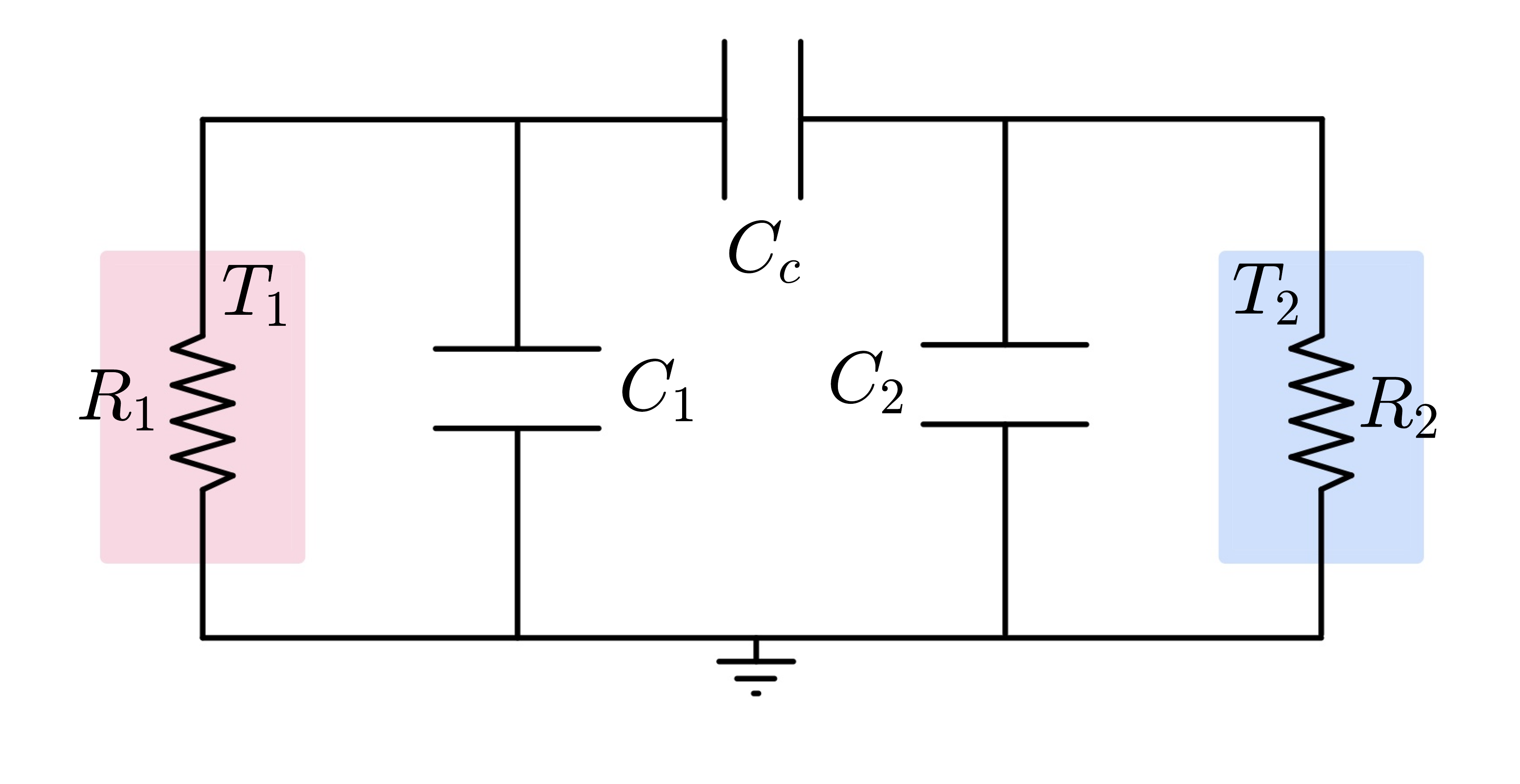}
\caption{Electrical embodiment of the Brownian gyrator~\cite{BGyrator2017electrical}.}
\label{fig:BG}
\end{figure}

We consider an electrical embodiment of a Brownian gyrator, see \cite{BGyrator2017electrical},  that consists of three capacitors and two resistors as shown in Fig.\ref{fig:BG}. The resistors are in contact with heat baths of different temperature.
The state of this electrical-thermal system comprises of the charges at two of the capacitors. Specifically, let $q_1(t)$ and $q_2(t)$ denote the charges at capacitances $C_1(t)$ and $C_2(t)$ (that are time-varying), and set
\begin{align*}
  &  q_t=\left[\begin{array}{c}
    q_1 (t) \\
     q_2 (t)
\end{array}\right],\qquad
R=\left[\begin{array}{cc}
     R_1& 0 \\
     0& R_2
\end{array}\right],
\end{align*} 
and $$
C(t)=\left[\begin{array}{cc}
    C_1(t)+C_c(t) &-C_c(t)  \\
    -C_c(t) & C_2(t)+C_c(t)
\end{array}\right],$$
for the charge vector, and the resistance and capacitance matrices, respectively. 
The dynamics are expressed in the  following two-dimensional stochastic differential equation,
\begin{equation}\label{eq:langevin-electric}
\ud q_t=-R^{-1}C^{-1}(t) q_t\ud t+R^{-1}DdB_t,
\end{equation}
where $\{B_t\}$ is a two-dimensional Brownian motion that models
 Johnson-Nyquist noise $R^{-1} D dB_t$ at the two resistors \cite{nyquist}, with
 \[
 D=\left[\begin{array}{cc}
    \sqrt{2k_BR_1T_1} & 0 \\
   0  &   \sqrt{2k_BR_2T_2} 
\end{array}\right],
\]
$k_B$ the Boltzmann constant, and $T_1$ and $T_2$ the temperature to which the corresponding resistors are subjected to. 

Denoting $U(t,q)=\frac{1}{2}q'C^{-1}(t)q$ the 
(potential) energy in the system of capacitances ($C_1(t),C_2(t),C_c(t)$),
the state equation \eqref{eq:langevin-electric} becomes
\begin{equation*}
\ud q_t=-R^{-1}\nabla_q U(t,q)\ud t+R^{-1}DdB_t,
\end{equation*}
where $\nabla_q$ is the gradient operator with respect the charge vector.
This is a two-dimensional overdamped Langevin equation, analogous to the equation that describes the motion of a particle with two-degrees of freedom, in a time-varying potential well, with $R$ playing the role of a viscosity matrix.

\eqref{eq:langevin-electric} represents a stochastic system whose state comprises of a probability distribution, denoted by  {$p(t,q)$}. This satisfies the Fokker-Planck equation
 \begin{equation*}\label{eq:Fokker-Planck}
    \frac{\partial p}{\partial t}  + \nabla_q \cdot J = 0,
\end{equation*}
with {probability current}
\begin{equation*}\label{eq:flux}
    J =\left[\begin{matrix}J_1\\J_2\end{matrix}\right]= -R^{-1}\left[\nabla_q U + {\frac{1}{2} DD'R^{-1}} \nabla_q \log(p)\right]p.
\end{equation*}

If the initial state {$p(t,q)$} is Gaussian with mean $0$ and covariance $\Sigma_0$, denoted by $N(0,\Sigma_0)$, then,
under time-varying quadratic potential (as in here), {$p(t,q)$} remains Gaussian for all times $t$,
with mean $0$ and
covariance $\Sigma(t)$ that satisfies the Lyapunov equation 
\begin{equation}\label{eq:Lyapunov}
    \dot\Sigma(t)=-R^{-1}C^{-1}(t)\Sigma(t)-\Sigma(t) C^{-1}(t)R^{-1}+R^{-1}DD' R^{-1}.
\end{equation}
The periodic variation of the capacitances that generates an attractive periodic orbit in the space of probability densities has been previously studied~\cite{EnergyHarvestingAnisotropic2021}. In that case, the periodic orbit is specified by a periodic covariance matrix for the charge vector $q_t$.

The coupling in \eqref{eq:langevin-electric} with the two heat baths allows transference of heat between the two, as well as exchange of energy with the environment through coupling with the time-varying potential $U$. Indeed, the total energy in the system (averaged over realizations) is
\begin{align*}
\mathcal E &= \mathbb E_{p}\{U(t,q)\}=\int Up\, dq=\frac{1}{2}\trace[C^{-1}(t)\Sigma(t)],
\end{align*}
where $\trace [ \cdot ]$ denotes trace. 
Likewise, the power delivered to the system via changes in the potential, is
\begin{align*}\nonumber
    \mathcal{ \dot{W}} &= \mathbb E_p\{\dot W\}=\int \frac{\partial U}{\partial t} p\, dq= \frac{1}{2}\trace[ \dot{C}^{-1}(t) \Sigma(t)],\label{eq:work-p}
\end{align*}
where
\[
\dot W=\frac{\partial U(t,q)}{\partial t}
\]
denotes the work rate along a single realization of the process.
The heat uptake from the respective thermal baths with temperature $T_1$ and $T_2$ is
\begin{align*}
    \mathcal{\dot{Q}}_k &= \int  J_k \partial_{q_k} U \, dq
    = - \int  U \partial_{q_k} J_k \,dq,
\end{align*}
for $k\in\{1,2\}$, respectively,
resulting in a total heat uptake
\begin{equation*}\label{eq:heat-p}
    \mathcal{\dot{Q}} = \mathcal{\dot{Q}}_1 + \mathcal{\dot{Q}}_2=- \int  U \nabla_q \cdot J\,dq = \frac{1}{2}\trace[ {C^{-1}}(t) \dot{\Sigma}(t)]. 
\end{equation*}
Note that $\frac{d}{dt}\mathcal E= \mathcal{ \dot{W}}+\mathcal{\dot{Q}}$,
in agreement with the first law, and that the time integrals of $\mathcal{ \dot{W}}$, $\mathcal{\dot{Q}}$ depend on the paths.

Basic physics dictates that the capacitance matrix $C(t)$ is positive definite at all times $t$. In the case were $C(t)=C_{\rm const}$ is constant, the covariance matrix specified in
\eqref{eq:Lyapunov} satisfies  $\Sigma(t)\to \Sigma_\infty$ as $t\to\infty$ and, thereby, the system
reaches a stationary steady state\footnote{ This follows from the fact that
the matrix $-(C_{\rm const} R)^{-1}$ is similar to $-(R^\frac12 C_{\rm const} R^\frac12)^{-1}$ which has negative eigenvalues, and as a consequence  the linear operator $\Sigma\mapsto -(C_{\rm const} R)^{-1}\Sigma-
\Sigma (C_{\rm const} R)^{-1}$ has negative spectrum as well. Hence, the limit $ \Sigma_\infty$ exists and satisfies the linear algebraic equation
$C_{\rm const} ^{-1}\Sigma_\infty R+R\Sigma_\infty C_{\rm const} ^{-1}=DD'$.}.
At steady state $\nabla \cdot J=0$, which implies vanishing total heat uptake. However, unless the detailed balance condition $J = 0$ is satisfied, the stationary steady state $N(0,\Sigma_\infty)$ is not an equilibrium distribution.  In such a stationary steady state, referred to as non-equilibrium steady state (NESS), the non-zero probability current mediates a heat transfer flux $\dot{\mathcal Q}_1=-\dot{\mathcal Q}_2\neq 0$ between the two thermal baths. Quantifying this heat flux has been the subject of earlier works~
\cite{BGyrator2007first,BGyrator2013CilibertoExperim,BGyrator2013dotsenko,BGyrator2017electrical,Imparato2017BG}.

In the sequel we are interested in the case where the capacitances $C_k(t)$ ($k\in\{c,1,2\}$) vary with time so as to allow extracting thermodynamic work out of the system. In our earlier work \cite{EnergyHarvestingAnisotropic2021} we quantified trade-offs between dissipation and work that can be extracted in similar Langevin systems via externally and periodically varying parameters. Herein, we pursue an alternative route where the relevant parameters (capacitances) are a function of added degrees of freedom, introduced
via coupling of the components, specifically the dielectric material in the capacitors, with an inertial wheel. This allows the autonomous function of the Brownian gyrator as a genuine autonomous thermodynamic engine.



\section{Engine concept}

\begin{figure}
\centering
\includegraphics[width=\linewidth, trim={0.5cm 0cm 0cm 0cm}, clip]{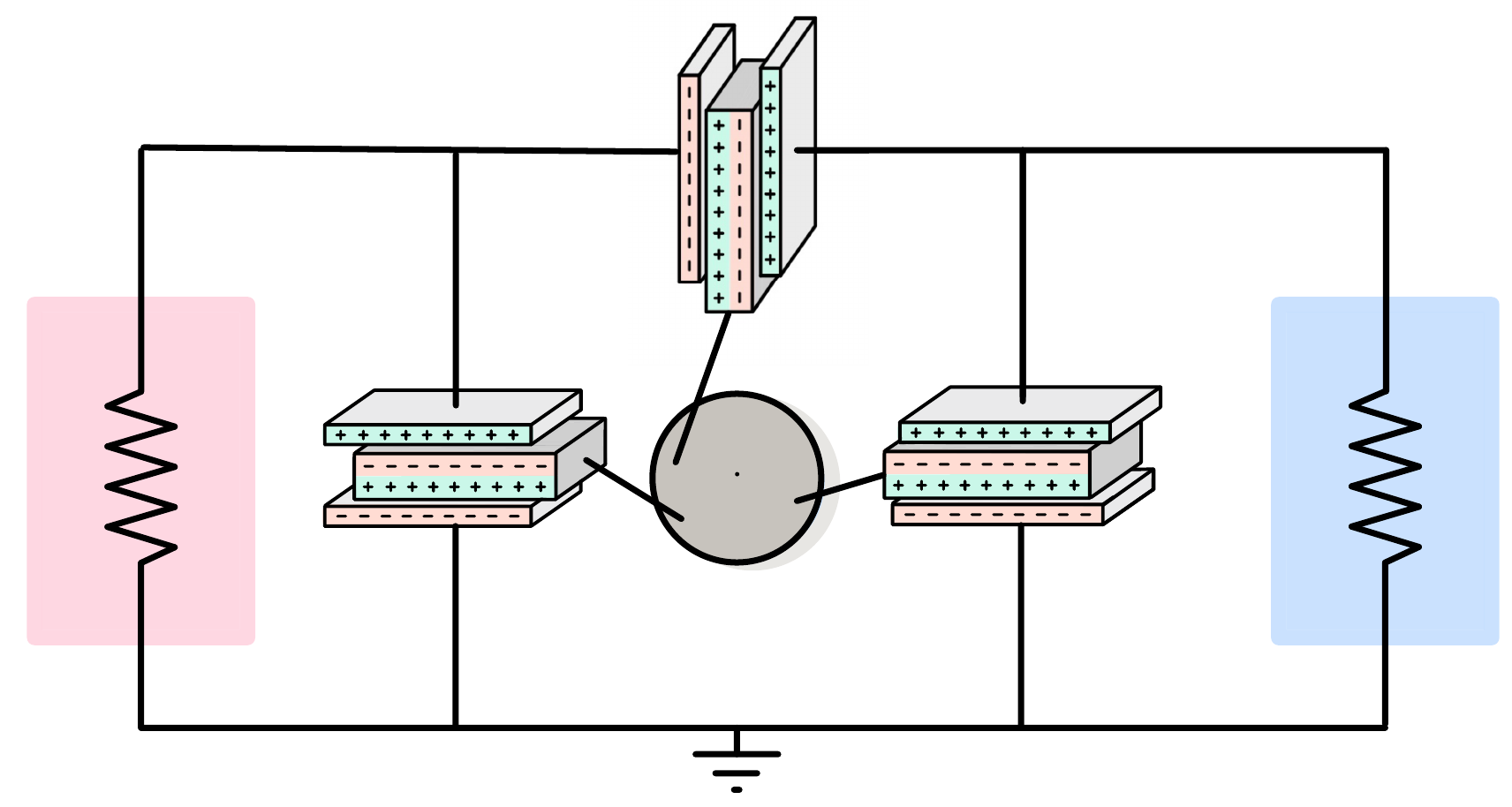}
\caption{Engine embodiment with actuated dielectric pads. The angular position $\theta_t$ of the wheel forces the dielectric material in and out of the capacitors, rendering the capacitance matrix a function of $\theta_t$. 
}
\label{fig:engine}
\end{figure}

Let us consider the presence of dielectric padding in the three capacitors, that can vary in position through mechanical coupling to a rotating wheel, as shown with a schematic in Fig.~\ref{fig:engine}. In this way, the angular position $\theta_t$ of the (inertial) wheel forces the dielectric material in and out of the respective capacitors. This mechanical coupling renders the capacitance matrix variable with time, being function of the dynamic variable $\theta_t$.

We select a geometry of the linkages actuating the dielectric material that gives the capacitance matrix as a function of $\theta$ in the form
\begin{equation*}
 \small   C(\theta)=\hspace{-1pt}C_0\hspace{-1pt}\left[ \begin{array}{cc}
         \hspace{-4pt}2+\beta g_1(\theta) &  {- 1-\beta \cos(\theta)} \\
       {-1-\beta\cos(\theta)} & \hspace{-1pt}2+\beta g_2(\theta)
    \end{array}\hspace{-4pt}\right]\hspace{-2pt},
\end{equation*}
where $g_1(\theta)=\cos(\theta+\phi_1) +\cos(\theta)$, $g_2(\theta)=\cos(\theta+\phi_2) +\cos(\theta)$ and $0<\beta<1$.
The specific form follows if we assume that i) the linkages are long enough so that the capacitances, to a good approximation, vary sinusoidally with the angular position $\theta$ of the wheel, and ii) that the links are attached to suitable positions to account for the phase differences $\phi_1,\phi_2$. 

The purpose of the mechanical coupling is to transfer the torque generated by the thermal fluctuations at the capacitors to the inertial wheel so as to average out, as well as provide needed phase difference (reflected in the parameters $\phi_1$, $\phi_2$) between the elements of the engine so as to sustain its continuous operation.

The model for the coupled system of the (electrical) Brownian gyrator, the inertial subsystem with linkages shown in Fig.~\ref{fig:engine}, and the external torque $-\gamma\omega_t$ to transfer work to the environment, is
\begin{subequations}\label{eq:motion-wheel}
\begin{align}
    \ud q_t&=-R^{-1}C^{-1}(\theta_t)q_t\ud t+R^{-1}DdB_t,\label{eq:qs}\\
    \ud \theta_t&=\phantom{-}\omega_t \ud t,\label{eq:theta-dot}\\
    I\ud \omega_t&=-\frac{1}{2}q_t'\partial_\theta C^{-1}(\theta_t)q_t\ud t-\gamma \omega_t \ud t. \label{eq:omega-dot}
\end{align}
\end{subequations}

In the above, the symbol $I$ represents the inertia of the wheel, and
$\gamma$ can be thought of as a friction coefficient in a process that helps extracting the work out of the engine. We will refer to the term $-\gamma\omega_t$ as external dissipation, though it could just as well represent torque proportional to $\omega$ exchanged with an external subsystem.

Noticing that
\[
\dot W=\frac{\partial U(t,q_t)}{\partial t}=
\frac{1}{2}q_t'\partial_\theta C^{-1}(\theta_t)q_t{\omega_t},
\]
we rewrite \eqref{eq:omega-dot},
\[
    I\omega_t \ud \omega_t=-\dot{ W}dt-\gamma \omega_t ^2\ud t,
\]
and integrate over time from $0$ to $t$ to obtain

\begin{equation*}
    \frac{1}{2}I(\omega_t ^2-\omega_0^2)= -W_t-\gamma \int_0^t\omega_s ^2\ud s,
\end{equation*}
where $W_t=\int_0^t \dot W ds$.
Therefore, the change in the kinetic energy of the wheel equals the work produced by the engine, minus the energy transferred via the torque $\gamma \omega_t$ to the environment (as friction or coupled to another system). Thus, it is intuitively clear that as long as the engine produces work, 
the parameter $\gamma$ can be adjusted to ensure that the wheel keeps rotating.


\section{Analysis}

We present an analysis that supports our claim and shows that, for a suitable set of parameters, the autonomous system generates positive work output over a cycle which sustains the rotational motion of the wheel and at the same time supplies torque to an external dissipative load. 

We adopt the assumption that there is a significant time-scale separation between the electrical and mechanical subcomponents of the engine, in that $\tau_{\rm elec}\ll\tau_{\rm mech}$, where 
$\tau_{\rm elec}$ and 
$\tau_{\rm mech}$ 
are the time-scales governing the electric and mechanical subsystems, respectively. 
With this time-scale separation, at every time-instant, $q_t$ can be viewed as a random vector which follows the stationary distribution of \eqref{eq:qs} associated with $\theta_t$. The distribution is Gaussian with zero mean and covariance matrix $\Sigma(\theta)$ that satisfies the algebraic Lyapunov equation
    \begin{equation}\label{eq:Lyapunoveq}
        -R^{-1}C^{-1}(\theta)\Sigma(\theta)-\Sigma(\theta) C^{-1}  (\theta)R^{-1}+R^{-1}DD' R^{-1} =0.
    \end{equation}
Moreover, the correlation of $q_t$ is localized over time due to its fast dynamics. Thus, the randomness it injects to \eqref{eq:omega-dot}  averages out fast so that, effectively, the mechanical subsystem is driven by the covariance $\Sigma(\theta)$ of the charge vector $q_t$. Hence, the dynamics of the mechanical components can be approximated by the deterministic dynamics
    \begin{subequations}\label{eq:deterministic}
    \begin{align}
    \dot\theta_t=&\phantom{+}\omega_t,\\
    I\dot \omega_t=&-\frac{1}{2}\trace[\partial_{\theta} C^{-1} (\theta)\Sigma(\theta)]-\gamma \omega_t.
    \end{align}
    \end{subequations}

The accuracy of the above approximation is positively correlated with the magnitude of the time-scale separation.
Elaborating further, the assumption on time-scale separation allows us to write $\theta$ and $\omega$ as functions of the time average of $q_tq_t'$ over $\tau_{\rm mech}$, i.e., $\frac{1}{\tau_{\rm mech}}\int_t^{t+\tau_{\rm mech}}q_sq_s'ds$. Since the stochastic system \eqref{eq:qs} is linear and the matrix $R^{-1}C^{-1}$ has positive eigenvalues, $q_t$ will approach, relatively fast, a steady state Gaussian distribution with variance $\Sigma(\theta)$. Thereby, we can replace this time average with $\Sigma(\theta)$.
\subsection{Phase portrait}
\begin{figure}
    \centering
    \includegraphics[width=0.4\textwidth,trim={6.8cm 14.75cm 4cm 6.4cm},clip]{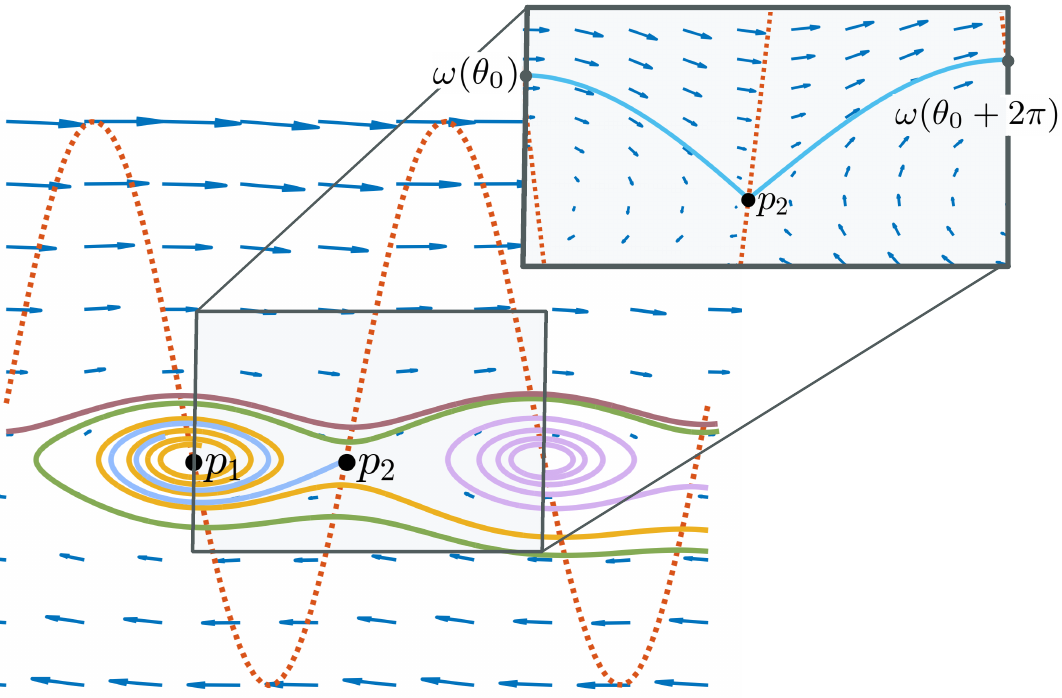}
    \caption{Schematic phase portrait of~\eqref{eq:deterministic}. The red dotted line marks the isoclene, while solid lines display different solutions in state space $(\theta,\omega)$. The point $p_1$ and $p_2$ label an stable and an unstable equilibrium point, respectively. In the blow up figure we have displayed a solution that crosses the neighborhood of the saddle point from above. The behaviour of this solution helps establish the existence of the stable periodic orbit. This periodic orbit is seen in the figure to ``weave'' at about $90^{\rm o}$-phase difference with the isoclene.}
    \label{fig:phaseportrait}
\end{figure}
The system of equations \eqref{eq:deterministic} describe motion in a force field,
\[
I\ddot \theta + \gamma \dot\theta = -F(\theta)
\]
where 
$
F(\theta)=
-\trace[\partial_{\theta} C^{-1} (\theta)\Sigma(\theta)]/2.
$
The dynamics are analogous to those
of a forced pendulum \cite{pendulumtorque}, and while in our case $F$ may not be exactly sinusoidal, qualitatively the response is quite similar.

A schematic of a phase plot, for a suitable choice of parameters, is shown in Fig.~\ref{fig:phaseportrait}. In this, the red dotted curve highlights the isoclene
\[
\omega=-\frac{1}{\gamma}F(\theta)
\]
which is periodic in $\theta$. The flow field above this curve points ``southeast,'' whereas below the curve, it points ``northeast.''
Points of equilibrium exist at the intersection of the isoclene with the $\omega=0$ axis, i.e., at points $(\theta,\omega=0)$ where $F(\theta)=0$. Over a $2\pi$-interval there are two such points of equilibrium. The first, labeled $p_1$ in Fig.~\ref{fig:phaseportrait}, is a stable focus, whereas the second, labeled $p_2$, is a saddle point (unstable). Their pattern repeats with a period equal to $2\pi$.

Only two types of trajectories are possible. Either the system settles to a oscillatory motion about a point of equilibrium, or it settles about a persistent periodic orbit where $\theta$ monotonically increases. Such a trajectory ``weaves'' with the isoclene and is highlighted in the figure (Fig.~\ref{fig:phaseportrait}) by a periodic solid line.

The existence of a stable periodic orbit requires sufficient work production over a cycle.
A sufficient condition that guarantees existence of such a stable periodic orbit can be deduced from the Poincare map $P$
that relates the value of $\omega(\theta)$ along orbits corresponding to values of $\theta$ that differ by $2\pi$:
\[
P: \omega(\theta_0) \mapsto \omega(\theta_0 +2\pi).
\]
To this end, we consider a trajectory that ``scrapes'' past the unstable equilibrium at $p_2$ as shown in the blowup of Fig. \ref{fig:phaseportrait}.
This trajectory can be numerically evaluated integrating forward and backwards in time, starting from a neighborhood of $p_2$. The inequality $\omega(\theta_0+2\pi)\geq \omega(\theta_0)$ guarantees the existence of a periodic orbit. 

In fact, if this periodic orbit exists, it must be unique, since there can only exist one periodic orbit that dissipates exactly the amount of work that is produced over a cycle. Specifically, work produced over a cycle is given by
$$
W_{\rm cycle}=-\frac{1}{2}\int_0^{2\pi}\trace[\partial_{\theta} C^{-1} (\theta)\Sigma(\theta)]\ud \theta,
$$
independently of the initial conditions. On the other hand, dissipation over a cycle is given by $\gamma \int_0^{2\pi}\omega(\theta) d\theta$, which is monotonic in the initial velocity $\omega(\theta_0)$. Therefore, if a periodic orbit exists, it is unique, since only one solution curve can satisfy $\gamma \int_0^{2\pi}\omega(\theta) d\theta=W_{\rm cycle}$.

In the following, we pursue an alternative route to ensure the existence of a periodic orbit. Our phase portrait analysis, however, helps assess numerically the existence of such an orbit.

\subsection{Persistence of motion}
\begin{figure}
    \centering
    \includegraphics[width=0.38\textwidth]{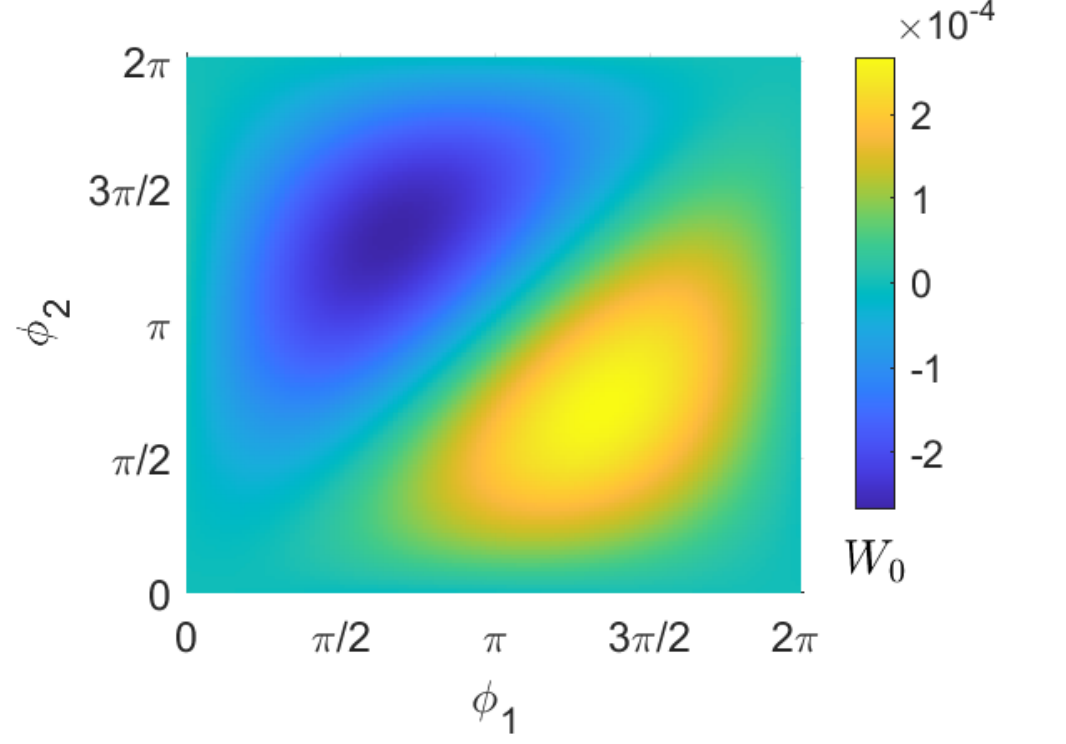}
    \caption{Dimensionless quasi-static work over a cycle, $W_0$, for different values of the phase differences $\phi_1$, $\phi_2$. Note that the maximum of $W_0$ is found at the point $(\phi_1^*,\phi_2^*)\approx(4.25,2.15)$.}
    \label{fig:heatmap-phis}
\end{figure}


To continue with our analysis, we express the solution $\Sigma(\theta)$
of \eqref{eq:Lyapunoveq} explicitly in terms of $\theta$ as follows,
\begin{align*}\label{eq:Sigmatheta}
\Sigma(\theta)&=\int_0^\infty e^{-R^{-1}C^{-1}(\theta)\tau}R^{-1}DD'R^{-1}e^{-R^{-1}C^{-1}(\theta)\tau}\ud \tau.
\end{align*}
Note that $\Sigma(\theta)$ is bounded for all $\theta$. 

In light of the structure of \eqref{eq:deterministic}, 
we may view $\theta$ as an independent variable (as long as $\omega>0$) and combine the two equations into a single equation that specifies the dependence of $\omega(\theta)$ on $\theta$. Considering a dimensionless velocity $\Omega=\gamma/(k_B T_1)\omega$, we can write
\begin{equation}
     \frac{d{\Omega}}{d\theta}(\theta)=\epsilon f(\theta,\Omega),\label{eq:omega(theta)}
\end{equation}
where $\epsilon=\gamma^2/(Ik_BT_1)$, and
$$f(\theta,\Omega)=-{\frac{1}{2k_BT_1}}\frac 1\Omega\trace[\partial_{\theta} C^{-1} (\theta)\Sigma(\theta)]-1$$
is a $2\pi$-periodic function of $\theta$. Note that $f(\theta,\Omega)$ is continuous and bounded, and so are its derivatives with respect to $\Omega$ up to second order on $(\theta,\Omega)\in[0,\infty)\times [\Omega_{\min},\Omega_{\max}]$
with  $\Omega_{\max}>\Omega_{\min}>0$. 
Therefore, we can infer stability from the averaged system as long as $\epsilon$ is small enough \cite[Theorem 10.4]{Khalil}, as explained below.

\begin{figure}[t]
    \centering
    \includegraphics[width=0.42\textwidth,trim={9.5cm 7.5cm 10.2cm 6cm},clip]{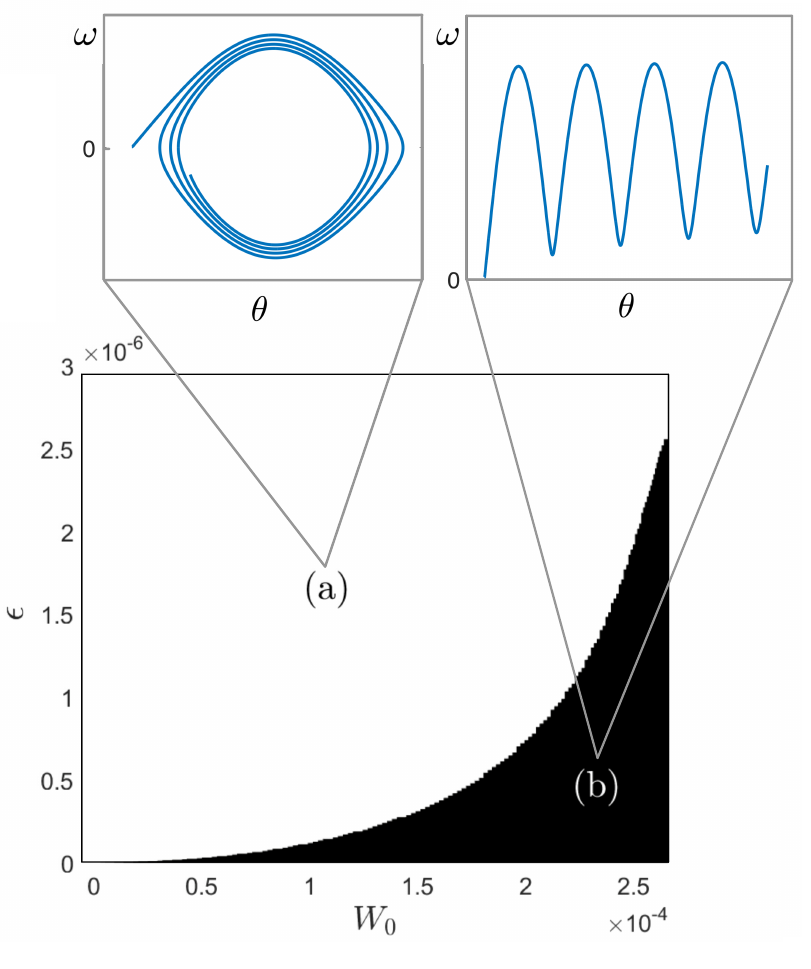}
   \caption{Space of parameters $(\epsilon, W_0)$ numerically divided into two regions (by testing weather  $\omega(\theta_0+2\pi)\geq \omega(\theta_0)$ is true or false): (a) parameters for which the only stable solution is the stationary solution, and (b) both the stationary solution and the periodic solution are stable. In the blow up figures we have, for each of the regions, an example trajectory in state-space.}
    \label{fig:bifurcation}
\end{figure}

Following \cite[Section 10.4]{Khalil}, define the {\em averaged system}
\begin{equation*}
    \frac{d{\bar\Omega}}{d\theta}(\theta)
    =\epsilon \Big(\frac{ W_0}{\bar\Omega}-1\Big),
\end{equation*}
where
\begin{align}\label{eq:avwork}
W_0&=\frac{W_{\rm cycle}}{2\pi k_BT_1},
\end{align}
is the dimensionless averaged work output over a complete rotation of the (deterministic) wheel.
It readily follows that $\bar\Omega=W_0$ is an exponentially attractive equilibrium for the averaged system.
Therefore, if $\bar \Omega(0)=W_0\in [\Omega_{\min},\Omega_{\max}]$ and $\Omega(0)-W_0=O(\epsilon)$, there exists $\bar\epsilon$ such that for all 
$\epsilon\in(0,\bar \epsilon)$, $\Omega(\theta)$ is defined and
\begin{equation*}
    \Omega(\theta)-W_0=O(\epsilon)\qquad \mbox{for all }\theta\in[0,\infty).
\end{equation*}
Moreover, $\Omega(\theta)$ is a unique, exponentially stable, $2\pi$-periodic solution of \eqref{eq:omega(theta)}.

Consequently, the wheel keeps rotating (i.e. $\Omega(\theta)>0$) provided that $W_0$, the dimensionless averaged work output in \eqref{eq:avwork}, dominates the $O(\epsilon)$ discrepancy.
Such a condition can be ensured by selecting the parameters $\phi_1$ and $\phi_2$ so that $W_0$ is positive, as can be seen in Fig.~\ref{fig:heatmap-phis}, and selecting $\epsilon$ sufficiently small. While, a precise bound on the size of $\epsilon$ as a function of $W_0$ is not available, it can be numerically determined as shown in Fig.~\ref{fig:bifurcation}. The figure divides the space of parameters $(\epsilon,W_0)$ into two regions, depending on the type of equilibrium obtained. In region (a) the wheel will eventually come to a stall, regardless of the initial conditions, since dissipation (characterized by $\epsilon$) dominates the work production. On the other hand, for parameters in region (b), the wheel will settle to a periodic orbit, as long as the system starts with sufficient momentum.  




\section{Numerical experimentation}

To demonstrate the validity of our results, we numerically compute realizations of the original process \eqref{eq:motion-wheel} and compare them to the solutions of the deterministic system~\eqref{eq:deterministic}. In order to do so, we ensure the time-scale separation by selecting our parameters such that $\sqrt{R_1R_2}C_0\ll\gamma/(k_BT_1)$ and $\sqrt{R_1R_2}C_0\ll I/\gamma$,
where $\sqrt{R_1R_2}C_0$ is the time-scale governing the electrical subsystem, while $\gamma/k_BT_1$ and $I/\gamma$ are the time-scales governing the oscillation and the damping, respectively, of the mechanical subsystem.

\begin{figure}
     \centering
    \includegraphics[width=0.48\textwidth, trim={2.15cm 19cm 11.5cm 2cm}, clip]{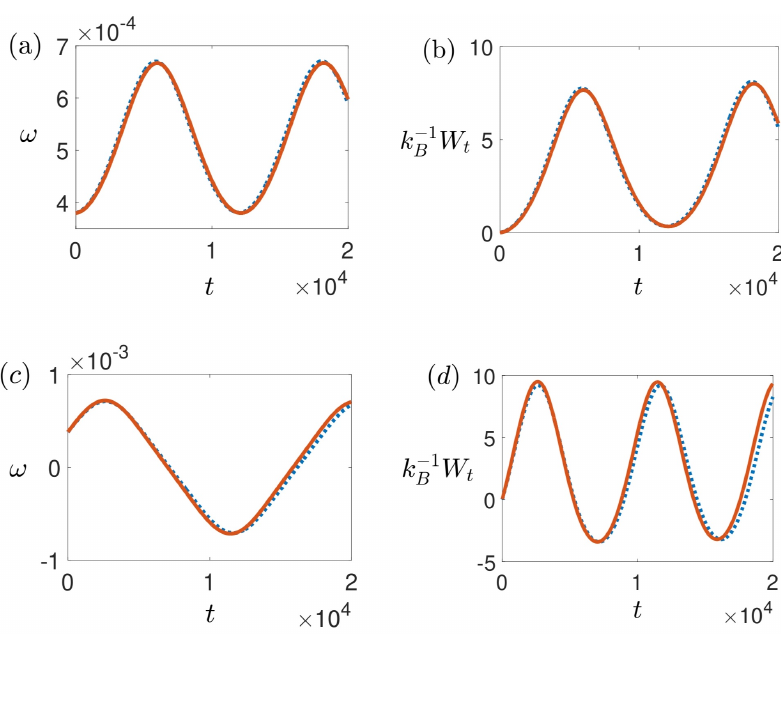}
        \caption{Solutions to the stochastic system \eqref{eq:motion-wheel} are displayed in solid red, while those to the deterministic system \eqref{eq:deterministic} are shown in dotted blue. Figures (a) and (b) are the result of a successful choice of parameters in that work output is produced and the wheel's motion is enhanced. On the other hand, in Figures (c) and (d) the work production is not  sufficient to overcome dissipation and thus the wheel oscillates around the static equilibrium point.}
        \label{fig:num-sols}
\end{figure}

Fig.~\ref{fig:num-sols} shows the solutions to the stochastic differential equations~\eqref{eq:motion-wheel} in solid red, while those to the deterministic equations~\eqref{eq:deterministic} are portrayed in dotted blue.   The top Figures (a and b) display a 
solution for which the work output is positive and increasing, thus augmenting the wheel's velocity (at least until it reaches the periodic orbit). On the other hand, Figures (c) and (d) showcase the opposite situation, the case where the wheel  eventually comes to a stall because the work output is not sufficient to overcome dissipation. To produce these plots, we used the parameters shown in Table~\ref{table}. All parameters were kept constant except for $\phi_1$ and $\phi_2$, which were varied between Figures (a) and (b) to Figures (c) and (d), from taking the optimal values to taking suboptimal ones.

We remark that although the stochastic solution in solid red constitutes only one realization, the time-scale separation ensures that all realizations behave approximately like the deterministic solution, and, therefore, typical solutions to the stochastic equation do not differ much from one realization to another. 

\section*{Final Remarks}
It is in order to point out the resemblance of the dynamical behaviour of our stochastic heat engine with that of a damped pendulum with constant torque \cite{pendulumtorque}. Indeed, due to the similarity of the shape of $F(\theta)$ to a shifted sine, it must be of no surprise that both systems share the same type of equilibria in the different regions of their parameter space. However, the damped pendulum \cite{pendulumtorque} has a qualitatively different behaviour when the constant torque applied is higher than the amplitude of undulations of the potential. In such a case, only the periodic solution is a stable solution.
We have not been able to replicate such a behaviour in our stochastic heat engine, which can be attributed to the fact that the amplitude of $F(\theta)$ is always greater than its vertical shift due to restrictions of the work production.

Besides the resemblance of the engine dynamics to those of a damped forced pendulum, there is also a qualitative resemblance to the dynamical behaviour of the Stirling engine as explained in \cite{StirlingTheoretical2018}. 
Specifically, a Stirling engine can be modeled as a periodic nonlinear pendulum, and its equilibrium modes have been experimentally shown to be same as those of our stochastic heat engine \cite{stirling2020}.
Thus, the analysis presented herein can be used to establish the existence of periodic orbits for the macroscopic Stirling engine, and to identify conditions for which periodic motion persists.
\begin{table}
\centering
\caption{Parameters in numerical simulation.}
\begin{tabular}{lrrr}
Parameter & Value & Units\\
\hline\\
 $I$ & 5$\times$ 10$^{2}$ & kg\,nm$^2$\\
 $\gamma$ & 10$^{-2}$ & \quad kg\,nm$^2$/s\\
 $\omega_0$ & 3.8$\times$10$^{-4} $& 1/s \\
 $\theta_0$ & $\pi/2$& rad\\
 $t_f$ & 2$\times $10$^4$& s \\
 $R_1$ & 1 & $\Omega$\\
 $R_2$ & 1 &  $\Omega$\\
 $T_1$ & 200 & K \\
 $T_2$ & 400 & K \\ 
 $k_B$ & 10 $^{-23}$ &  J/K\\
 $C_0$ & 2 & mF\\
 $\beta$ & 0.1 & -\\
 $\phi_1$& (a)$\&$(b): 4.25\ \ /\ \   (c)$\&$(d): $\pi$& rad\\
  $\phi_2$& (a)$\&$(b): 2.15\ \ /\ \ (c)$\&$(d): $\pi$& rad\\[2pt]
\hline
\end{tabular}
\label{table}
\end{table}

{\em Acknowledgements.--}
The project received the support of a fellowship from ”la Caixa” Foundation (ID 100010434) with code is LCF/BQ/AA20/11820047.
Additional support from NSF under grants
1839441, 1901599, 1942523 and AFOSR under grant FA9550-20-1-0029 are acknowledged.

\bibliography{arXiv2}

\end{document}